\documentclass{elsart}
\usepackage{graphicx,amssymb,lineno,color}

\begin{document}

\begin{frontmatter}
\title{Hydrogen adsorption on hexagonal silicon nanotubes}

\author[Sejong]{Junga Ryou},
\author[Sejong]{Suklyun Hong\corauthref{cor}}
\ead{hong@sejong.ac.kr}
\author[SKKU]{Gunn Kim\corauthref{cor}}
\ead{kimgunn@skku.edu}
\corauth[cor]{Corresponding authors. Tel.:
+82 2 3408 3209; fax: +82 2 497 2634}

\address[Sejong]{Department of Physics and Institute of Fundamental Physics,
Sejong University, Seoul 143-747, Korea}
\address[SKKU]{BK21 Physics Research Division and Department of Physics,
Sungkyunkwan University, Suwon 440-746, Korea}

\begin{abstract}
We present a first-principles study of geometrical structure and energetics
of hydrogen adsorbed on hexagonal single-walled silicon
nanotubes (SiNTs). The adsorption behaviors of hydrogen molecules in
SiNTs are investigated. The binding energies for the most stable
physisorbed configurations are calculated to be less than 0.1 eV.
The energy barriers are also investigated for dissociation of H$_2$
molecules. Finally, we consider encapsulation of H$_2$ molecules in
SiNTs. The possibility of SiNTs as hydrogen storage materials is
discussed.
\end{abstract}

\end{frontmatter}

\section{Introduction}
\label{intro}

Due to their novel quantum mechanical phenomena as well as various
potential applications, carbon-based low-dimensional nanostructures
such as carbon nanotubes and graphene have attracted much attention.
In fact, the recent synthesis of graphene provided foundation for
new devices. Nevertheless, silicon is still one of the most
important materials even in the field of nanoscience. Si-based
nanowires and nanotubes have recently been investigated
theoretically and
experimentally~\cite{Fagan1999,Barnard2003,Bai2004,Zhang2005,Durgun2005}.

Bai {\em et al.}~\cite{Bai2004} suggested that Si nanotubes (SiNTs)
could form by the top-to-top stacking of square, pentagonal and
hexagonal silicon structures and showed using ab initio calculations
that the pentagonal and hexagonal SiNTs may be locally stable in
vacuum. They provided a computational evidence for existence of such
one-dimensional silicon nanostructures, although those structures
have not been observed experimentally yet. Very recently, the
stability and related electronic structure of their hexagonal SiNTs
with some defects were investigated~\cite{Kim2008}.

Under ambient conditions, Si nanostructures may be passivated or
oxidized. To consider the possibility of SiNTs as a hydrogen storage
material along with hydrogen passivation, in this paper, we focus on
the interaction of hydrogen with pure SiNTs without defects. Here,
we study adsorption behaviors of hydrogen molecules on and in the
SiNT. The binding energies and the energy barriers are also
calculated.

\section{Calculation Methods}
\label{cal}

We have performed the density functional theory (DFT) calculations
within generalized gradient approximation (GGA) and local density
approximation (LDA) using the Vienna ab initio simulation package
(VASP)~\cite{Kresse1993,Kresse1996}. The cutoff energy is set to 200
eV and the atoms are represented by ultrasoft pseudopotentials as
provided with VASP~\cite{Kresse1994}. To investigate a hydrogen
molecule adsorbed on the SiNT, we consider hexagonal SiNTs having 48
silicon atoms and vacuum region between SiNTs about 10~\AA\
perpendicular to the tube axis. For the Brillouin-zone integration
we use a 1$\times$1$\times$15 grid in the Monkhorst-Pack special
k-point scheme. To study the reaction energy barriers for
dissociation and encapsulation of a hydrogen molecule on the SiNT,
we use the nudged elastic band (NEB) method~\cite{Henkelman2000}.

\section{Results and Discussion}
\label{res}

\begin{figure}
\begin{center}
\includegraphics*[width=10cm]{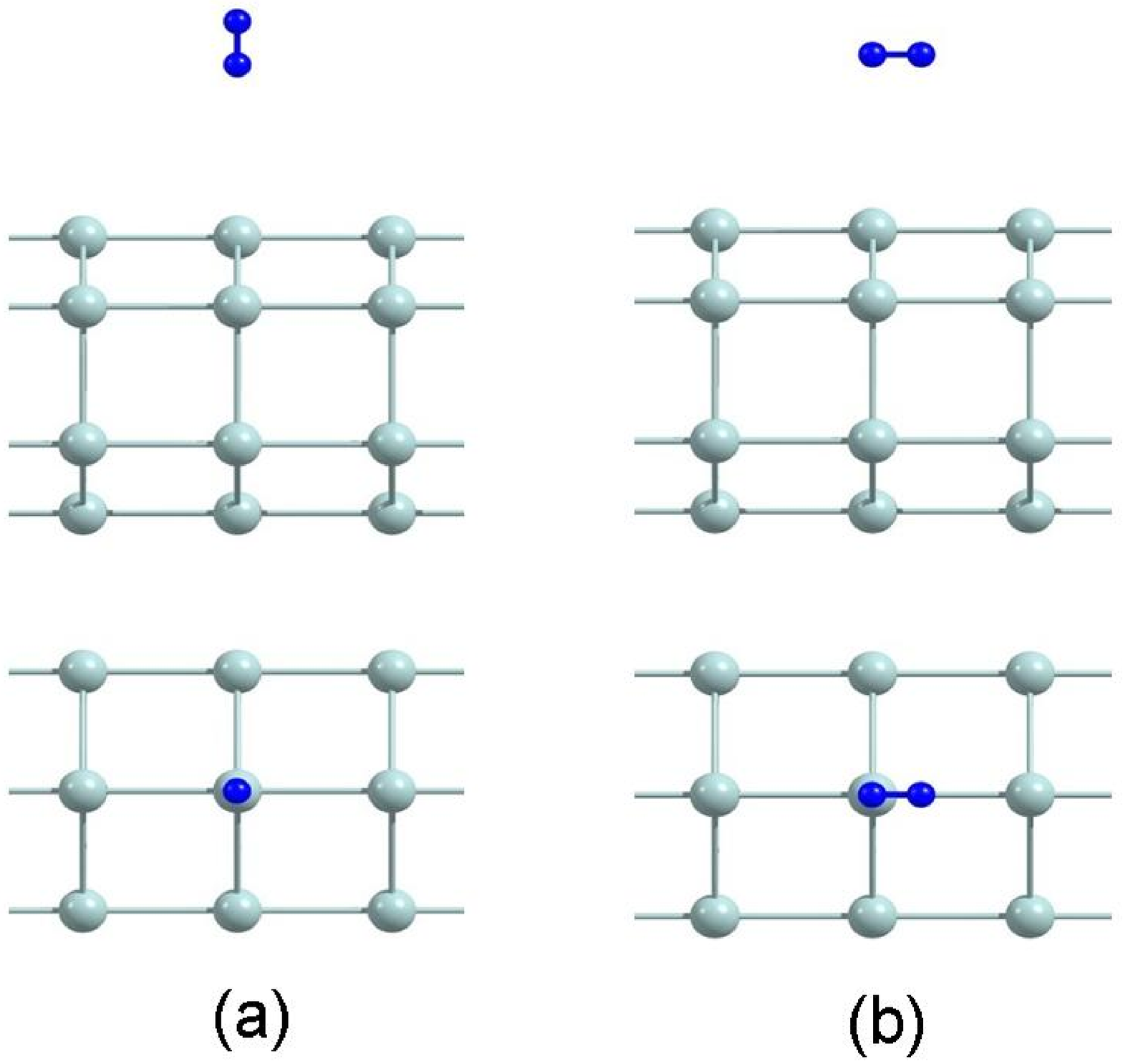}
\end{center}
\caption{Structure of hexagonal SiNTs with H$_2$ molecules adsorbed
on the atop site where the H-H bond is (a) perpendicular and (b)
parallel to the tube axis.} \label{fig:atop}
\end{figure}

First, we calculate the total energy of a SiNT with a hydrogen molecule
adsorbed at both the GGA and LDA levels. To find the most stable
configuration, several adsorption sites are considered, depending on
the position and orientation of H-H bond. The molecule can be
located at the top of Si atom (atop), the bridge of Si-Si bond
(bridge), and the center of Si square (hollow), respectively. For
each adsorption type, the H-H bond can be perpendicular or parallel
to the tube axis. Figure~\ref{fig:atop} shows two possible
configurations of the atop site.

The relative energies of adsorption sites are shown in
Table~\ref{tab:bindene}. In GGA, the atop site with perpendicular
H-H bond has the lowest energy (binding energy = 31 meV/H$_2$),
which is almost similar in energy to other perpendicular
configurations. In fact, the energy variations are very small, so
all configurations considered are within less than 10 meV in energy.
On the other hand, in LDA, the energy variations are relatively
large compared to the GGA case. The lowest energy configuration for
LDA is that H$_2$ is at the hollow site with perpendicular H-H bond,
and its binding energy is 94 meV/H$_2$. The distance between H$_2$
and the closest Si atom of SiNT is about 3.7~\AA\ (GGA) or about
3~\AA\ (LDA) for both the atop and hollow sites with perpendicular
H-H bond. Calculated binding energies and bond lengths manifest that
the LDA overbinds compared to the GGA.

\begin{table}
\caption{Relative energies (in units of meV/molecule) of adsorption
sites. The energies are referred to the lowest energy in each case
of GGA and LDA.}\label{tab:bindene}
\begin{tabular}{|c|c|c|c|c|c|c|}
\hline site & \multicolumn{2}{|c|}{atop} &
\multicolumn{2}{|c|}{bridge} & \multicolumn{2}{|c|}{hollow} \\
\hline &perp. & parallel &perp. & parallel &perp. & parallel \\
\hline GGA & 0  & 4.2& 0.6 &2.3& 0.6
&5.0 \\
\hline LDA &35.4&    51.4 &   4.5& 15.3 &   0 &  15.7\\
\hline
\end{tabular}
\end{table}

\begin{figure}
\begin{center}
\includegraphics*[width=6cm,angle=-90]{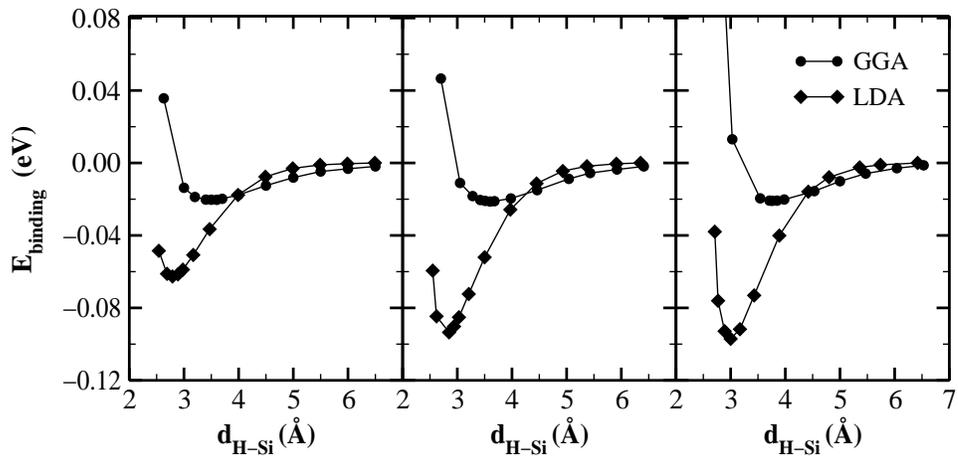}
\end{center}
\caption{Binding energy profile of H$_2$ adsorbed on SiNTs at (a)
atop site, (b) bridge site, and (c) hollow site, as a function of
the Si-H distance.} \label{fig:bindene}
\end{figure}

To better understand the binding behavior, we move the hydrogen
molecule toward the relatively stable site, i.e., the perpendicular
case of each adsorption type. Figure~\ref{fig:bindene} shows the
binding energy of the hydrogen molecule as a function of its
distance from the SiNT wall. In the calculation, the position of the
hydrogen molecule is fixed, while those of all Si atoms in the SiNT
are relaxed.
In GGA, the binding energy of hydrogen is about 0.02 eV and the
binding distance between the hydrogen molecule and a Si atom is
around 3.6-3.7~\AA\ for all three cases shown in
Fig.~\ref{fig:bindene}. On the other hand, in LDA, the binding
energy and distance are somewhat varied: the binding energy is 0.06
eV for the atop site and about 0.1 eV for other two cases, and the
binding distance is 2.8~\AA\ for the atop and bridge sites and
3.0~\AA\ for the hollow site. Since the position of the hydrogen
molecule is fixed, the (LDA) binding energies from
Fig.~\ref{fig:bindene} are slightly smaller within $\sim$0.01 eV
than the fully relaxed ones.

It is worthwhile to consider the possibility of SiNTs as hydrogen
storage materials. Many studies of hydrogen-adsorbed nanostructures
show that the binding energy for hydrogen storage should be about
0.3 eV~\cite{Lee2007,Lochan2006,Hamaed2008}. However, the hydrogen
molecule adsorbed on the SiNTs has a binding energy of 0.03 eV for
GGA and at most 0.1 eV for LDA, respectively. Previous studies
mentioned that the weak intermolecular (or nonbonding) bindings may
be similar to the LDA results or some halfway between the LDA and
GGA results~\cite{Sung2007}. Hence, the maximum value of the binding
energy of H$_2$ in the present system may be almost as large as 0.1
eV. Nevertheless, this binding energy is not enough for good
hydrogen storage materials. On the other hand, it is understood that
the weak nonbonding interaction (e.g., the van der Waals
interaction) such as physisorption might not be well described by
GGA or LDA~\cite{Kamiya2002,Millet1999,Wu2002}. Also, the van der
Waals interaction due to polarization in metallic systems might be
more important than in insulating or semiconducting systems.
However, H$_2$ has no electric dipole moment but has the electric
quadrupole moment, so the polarization effect is expected to be very
small for the H$_2$ adsorption. Recently, Henwood and Carey reported
that the binding energies of H$_2$ on the metallic (9,0) and the
semiconducting (10,0) CNTs are almost similar at several adsorption
sites for LDA and GGA, respectively~\cite{Henwood2007}. These CNT
results may imply that the nonbonding interaction due to
polarization in metallic SiNTs does not change our conclusion.
Therefore, even taking both inaccuracy of GGA or LDA for the
nonbonding interaction and the polarization effect into account, we
believe that the binding energy cannot be larger than 0.2-0.3 eV,
the required mininum binding energy for hydrogen storage.

\begin{figure}
\begin{center}
\includegraphics*[width=10cm]{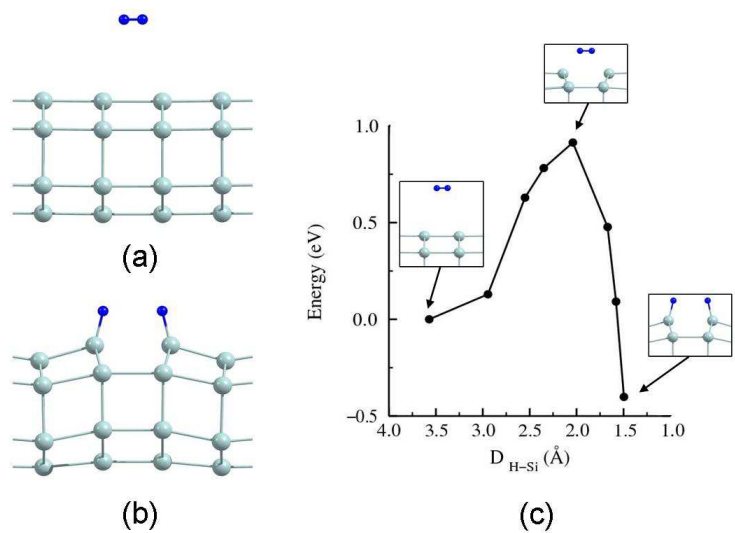}
\end{center}
\caption{Dissociation of H$_2$: (a) initial physisorbed state, (b)
final dissociated state, and (c) the reaction pathway, as a function
of the Si-H distance, from physisorbed to chemisorbed state showing
a dissociation barrier ($\sim$ 0.9 eV).} \label{fig:disso}
\end{figure}

We have considered the adsorption behaviors of the hydrogen molecule by
varying its initial distance from the SiNT wall. In some specific
cases (using GGA) where its bond is parallel to the tube axis and
its distance is quite close ($\sim$1.7~\AA) to the wall, the
hydrogen molecule at the bridge site is dissociated into hydrogen
atoms that are chemisorbed on nearby Si atoms. To find the activation energy
barrier of the dissociation pathway from a physisorbed state
[Fig.~\ref{fig:disso}(a)] to this chemisorbed state
[Fig.~\ref{fig:disso}(b)], we use the NEB
method~\cite{Henkelman2000}. We find that the bond of Si atoms with
hydrogen attached is broken in the chemisorbed state.
Figure~\ref{fig:disso}(c) shows that the energy barrier for
dissociation is about 0.9 eV. The energy barrier for hydrogen
dissociation on the SiNT is found to be lower than that on carbon
nanotubes (CNTs): For the (5,5) CNT case, the energy barrier is 2.70
or 3.07 eV for two reaction paths~\cite{Lee2005}, while that for the
(3,3) CNT is 2.7 eV~\cite{Chen2005}. On the other hand, a very small energy
barrier of about 0.3 eV was also reported~\cite{Tada2001}.

\begin{figure}
\begin{center}
\includegraphics*[width=10cm]{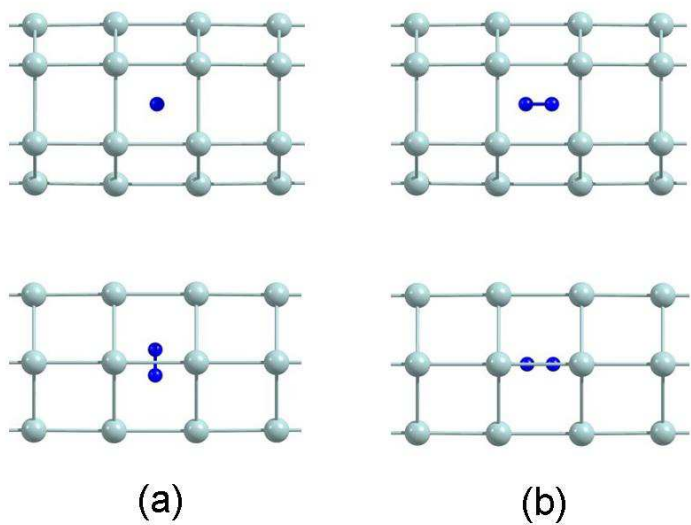}
\end{center}
\caption{Hydrogen molecule encapsulated in SiNTs, where the H-H bond
is (a) perpendicular and (b) parallel to the tube axis.}
\label{fig:encap}
\end{figure}

Finally, we consider a hydrogen molecule encapsulated in a SiNT.
Figure~\ref{fig:encap} shows that the encapsulated hydrogen molecule
is located at the center of the tube without dissociated.
Encapsulation energy is defined as $E$(encapsulation) =
[$E$(SiNT)+$E$(H$_2$)]-$E$(H$_2$@SiNT), where $E$(SiNT), $E$(H$_2$),
and $E$(H$_2$@SiNT) are the total energy of SiNT, H$_2$, and
H$_2$@SiNT, respectively. The H$_2$ encapsulation energy is
calculated to be -1.28 eV in both configurations shown in
Fig.~\ref{fig:encap}. The negative value means that the encapsulated
state is higher in energy than the separate configuration of H$_2$
and SiNT, and then H$_2$ is hard to be encapsulated in SiNTs
unless located in SiNTs by constraint. This encapsulation energy is
larger by 1.31 eV than the adsorption energy of the most stable
configuration on SiNTs. We consider pathways for penetration of
the hydrogen molecule into the tip-opened SiNT. Using the NEB
method~\cite{Henkelman2000}, the energy barrier is calculated to be
about 1.28 eV. Previous calculations shows that the penetration
barrier into the open end of the (3,3) CNT is about 3
eV~\cite{Tada2001}. This relatively larger value in the case of CNT
compared to SiNT may be due to difference in bond strength of C-H
and Si-H.

\section{Conclusion}
\label{concl}

We have investigated a hydrogen molecule adsorbed on SiNTs using DFT
calculations. Our calculations show that H$_2$ is the most stable at
the atop site (GGA) or the hollow site (LDA) with the H-H bond
perpendicular to the tube axis. However, since the binding energy
variations are very small, at low temperature, H$_2$ can adsorb
practically at any possible sites considered. For the most stable
physisorbed configurations, the binding energies are calculated to
be less than 0.1 eV. Thus, it is found that pure SiNTs are not good
for hydrogen storage. In addition, the activation energy barrier is
calculated to be about 0.9 eV for dissociation of H$_2$.
H$_2$ is not likely to be encapsulated in SiNTs.

\section*{Acknowledgement}

This research was performed for the Hydrogen Energy R\&D Center, one
of the 21st Century Frontier R\&D Program, funded by the Ministry of
Science and Technology of Korea. This work was also supported by the
grant from the KOSEF through the Center for Nanotubes and
Nanostructured Composites (CNNC) and by the second BK21 project of
Ministry of Education.

\clearpage

\end{document}